\documentclass[11pt,a4paper]{article}
\pdfoutput=1
\usepackage{jheppub}

\usepackage{multirow, graphicx,amssymb,url,mathrsfs,amsmath}
\usepackage{wrapfig,boxedminipage,setspace,subfigure,epsfig}
\usepackage{amsxtra,amstext,latexsym,dsfont,amsfonts}
\usepackage{color,eucal}
\usepackage[dvipsnames]{xcolor}
\usepackage{float}
\usepackage{slashed,comment}
\usepackage{kotex}

\usepackage{epigraph}

\usepackage[normalem]{ulem}

\newcounter{defcounter}
\usepackage[utf8]{inputenc}
\input Starburst.fd
\newcommand*\initfamily{\usefont{U}{Starburst}{xl}{n}}\initfamily

\newcommand{\beq}{\begin{eqnarray}}
\newcommand{\eeq}{\end{eqnarray}}
\usepackage{amsmath}
\usepackage{tikz}
\usetikzlibrary{decorations.pathmorphing}
\usetikzlibrary{shapes.misc}
\tikzset{cross/.style={cross out, draw=black, minimum size=8*(#1-\pgflinewidth), inner sep=0pt, outer sep=0pt},
cross/.default={1pt}}
\usetikzlibrary{patterns,math}

\title{On the universality of AdS$_2$ diffusion bounds and the breakdown of linearized hydrodynamics}

\author[a]{Ning Wu,}
\author[b,c]{Matteo Baggioli,}
\author[a]{Wei-Jia Li}
\emailAdd{b.matteo@sjtu.edu.cn}
\emailAdd{weijiali@dlut.edu.cn}
\affiliation[a]{Institute of Theoretical Physics, School of Physics, Dalian University of Technology,
Dalian 116024, China.}
\affiliation[b]{Wilczek Quantum Center, School of Physics and Astronomy, Shanghai Jiao Tong University, Shanghai 200240, China.}
\affiliation[c]{Shanghai Research Center for Quantum Sciences, Shanghai 201315.}

\abstract{The chase of universal bounds on diffusivities in strongly coupled systems and holographic models has a long track record. The identification of a universal velocity scale, independent of the presence of well-defined quasiparticle excitations, is one of the major challenges of this program. A recent analysis, valid for emergent IR fixed points exhibiting local quantum criticality, and dual to IR AdS$_2$ geometries, suggests to identify such a velocity using the time and length scales at which hydrodynamics breaks down -- the equilibration velocity. The latter relates to the radius of convergence of the hydrodynamic expansion and it is extracted from a collision between a hydrodynamic diffusive mode and a non-hydrodynamic mode associated to the IR AdS$_2$ region. In this short note, we confirm this picture for holographic systems displaying the spontaneous breaking of translational invariance. Moreover, we find that, at zero temperature, the lower bound set by quantum chaos and the upper one defined by causality and hydrodynamics exactly coincide, determining uniquely the diffusion constant. Finally, we comment on the meaning and universality of this newly proposed prescription.}

\begin{document}
\maketitle

\section{Introduction}
 \epigraph{The more specific we are, the more universal something can become.}{Jaqueline Woodson}
 The search for universal features in the transport properties of many-body quantum systems, strongly coupled materials and holographic models has a long history. In this context, universality is intended as insensitivity to the specific microscopic details of the system and it therefore resonates nicely with the concept of hydrodynamics \cite{landau2013fluid}. Hydrodynamics is an effective description controlling the long time and large scales dynamics, where all the short-lived operators carrying the microscopic information get washed out. In this sense, the universal properties remaining are carried by the long-lived quantities and they are consequently related to the so-called hydrodynamic modes and the corresponding conservation equations. Importantly, using this broad terminology, hydrodynamics can be applied to any physical systems and is not restricted to the description of fluids \cite{PhysRevA.6.2401}.\\
 
 In this ballpark, a milestone result has been the identification of a universal lower bound on the ratio of shear viscosity $\eta$ to entropy density $s$, which supposedly holds for any system in nature. The resulting inequality
 \begin{equation}
     \frac{\eta}{s}\,\geq\,\frac{\hbar}{4\,\pi\,k_B}\label{KSS}
 \end{equation}
 takes the name of Kovtun-Son-Starinets (KSS) bound \cite{Policastro:2001yc} and it has been derived using a dual gravitational description in terms of the black hole horizon dynamics and the gravitons absorption rate therein \cite{Gubser:1997yh,Klebanov:1997kc}. Notice how this bound immediately connects with hydrodynamics since the $\eta/s$ ratio coincides exactly with the transverse momentum dimensionless diffusion constant in a neutral relativistic fluid \cite{Kovtun:2012rj}. Indeed, for neutral relativistic systems, the KSS bound can be re-written as:
 \begin{equation}
     D_{\text{shear}}\,\geq\,\frac{c^2}{4\,\pi}\,\tau_{pl} \label{kss2}
 \end{equation}
 where $D_{\text{shear}}$ is the diffusion constant of the shear mode, $c$ the restored speed of light and $\tau_{pl}\equiv \hbar/k_B T$ the so-called Planckian time \cite{Zaanen2004,10.21468/SciPostPhys.6.5.061}. This last quantity plays an important role and it has been involved in several discussions and experimental observations about universality and transport \cite{Bruin804,Behnia_2019,PhysRevX.5.041025,PhysRevLett.123.066601,Policastro:2001yc,Maldacena:2015waa,Mousatov2020,PhysRevLett.124.076801,PhysRevLett.120.125901,Zhang19869,Lucas:2018wsc,Hartman:2017hhp}\\
 
 Despite the great success of the KSS bound even when confronted with realistic experimental data \cite{Schafer:2009dj,Cremonini:2011iq,Luzum:2008cw,Nagle:2011uz,Shen:2011eg}, it soon became clear that the inequality in Eq.\eqref{KSS} could be violated by breaking explicitly and/or spontaneously spacetime symmetries, such as translations and rotations\footnote{Curiously, this is slightly imprecise for the case of rotations. Indeed, contrary to the explicit breaking scenario, the spontaneous breaking of rotations does not imply a violation of the KSS bound \cite{ERDMENGER2011301}}. The references reporting on these violations are indeed several \cite{Alberte:2016xja,Hartnoll:2016tri,Burikham:2016roo,Rebhan:2011vd,Ge:2018lzo,Gochan:2018eez,Figueroa:2020tya}. From the physical point of view, these cases were accompanied by the observation that the ratio $\eta/s$ plays a very special role only in relativistic neutral fluids, while it is not connected with any specific transport properties otherwise. A clear example is that of a non-relativistic system in which the KSS bound can be violated just by increasing the number of different species \cite{PhysRevLett.99.021602}. In view of these facts, the universal character of the KSS bound has been recently discredited \cite{Baggioli:2020ljz}.\\
 
  In a parallel line of investigation \cite{Trachenkoeaba3747,Baggioli:2020lcf,Trachenko:2020jgr}, the superior (in the sense of more general and universal) role of the diffusion constants  (compared to the $\eta/s$ ratio for example) has been outlined in the context of realistic liquids and simple bounds on momentum and energy diffusion have been derived in terms of few fundamental physical constants.\\
 
 Inspired by the equivalent formulation of the KSS bound in terms of the shear diffusion constant expressed in Eq.\eqref{kss2}, a more general universal bound was later proposed in \cite{Hartnoll:2014lpa}. The idea is that any diffusive process in nature is bounded from below by a certain combination of two unknown velocity and time scales as:
 \begin{equation}
     D\,\geq\,v^2_?\,\tau_?\,. \label{hartnollbound}
 \end{equation}
 This last expression recovers immediately the KSS bound by setting $D=D_{\text{shear}}$, $v_?=c$ and $\tau_?=\tau_{pl}$.\\
 The inequality in Eq.\eqref{hartnollbound} applies to physical diffusion constants, it is very general and it can be consistently defined for any system possessing a diffusive hydrodynamic process which corresponds to the time evolution of a certain conserved quantity. Nevertheless, it appears quite void and not practical unless one specifies in detail which are the scales appearing in the r.h.s. of Eq.\eqref{hartnollbound}. Additionally, one would call such an expression universal only if the same velocity and time scales bounded all the diffusive processes in the system. On the contrary, a statement like Eq.\eqref{hartnollbound} would become quite poor if, for any diffusion constant $D_i$, different scales in the r.h.s. had to be used. Finally, following this logic, one would expect the scales in the r.h.s. of Eq.\eqref{hartnollbound} to be infrared (IR) quantities, independent of the ultraviolet (UV) microscopic physics, and therefore universal.\\
 
 A first, and partially successful, attempt to make the bound in Eq.\eqref{hartnollbound} more concrete has originated from the idea of identifying the scales in the r.h.s. using physical observables from quantum chaos. In particular, Refs. \cite{Blake:2016wvh,Blake:2016sud} have proposed to identify:
 \begin{equation}
     v_?\,=\,v_B\,,\qquad \tau_?\,=\,\tau_L,
 \end{equation}
 where $v_B$ is the butterfly velocity and $\tau_L$ the Liapunov time. Both these quantities can be directly extracted using the out-of-time-order correlator (OTOC) \cite{1969JETP...28.1200L}. In summary, the final proposal coming from \cite{Blake:2016wvh,Blake:2016sud} was that any diffusion constant $D_i$ has to be bounded from below as follows:
 \begin{equation}
     D_i\,\geq\,\#_i\,v_B^2\,\tau_L \label{blakebound},
 \end{equation}
 where $\#_i$ is an $\mathcal{O}(1)$ number which depends on the specific diffusive process as well as IR fixed point considered.\\
 Despite the considerable success of this proposal \cite{Davison:2018ofp,Gu:2017njx,Ling:2017jik,Gu:2017ohj,Blake:2016jnn,Blake:2017qgd,Wu:2017mdl,Li:2019bgc,Ge:2017fix,Li:2017nxh,Ahn:2017kvc,Baggioli:2017ojd,Kim:2017dgz,Aleiner:2016eni,Patel:2017vfp,Patel:2016wdy,Bohrdt:2016vhv,2017arXiv170507895W,Kim:2017dgz,Ahn:2017kvc,Chen:2020bvf,Lucas:2018wsc}, it was soon realized that in the case of charge diffusion $D_i\equiv D_Q=\sigma/\chi_{\rho\rho}$ (with $\sigma$ the electric conductivity and $\chi_{\rho\rho}$ the charge susceptibility) the bound in Eq.\eqref{blakebound} could be violated \cite{Lucas1608,Baggioli:2016pia}. One more time, this is not surprising from a physical point of view. In fact, the quantum chaos data ($v_B$ and $\lambda_L$) are extracted holographically from the gravitational sector of fluctuations and in general are totally agnostic about the charge sector to which the charge diffusion constant attains. When the diffusive process considered is that of energy, $D_i\equiv D_\epsilon=\kappa/c_v$ (with $\kappa$ the thermal conductivity and $c_v$ the specific heat), the bound in Eq.\eqref{blakebound} is much more robust and hard to break. Nevertheless, there are at least two known cases \cite{Lucas:2016yfl,wu2021classical} where this happens.\\
 \begin{figure}[ht]
     \centering
     \includegraphics[width=0.6\linewidth]{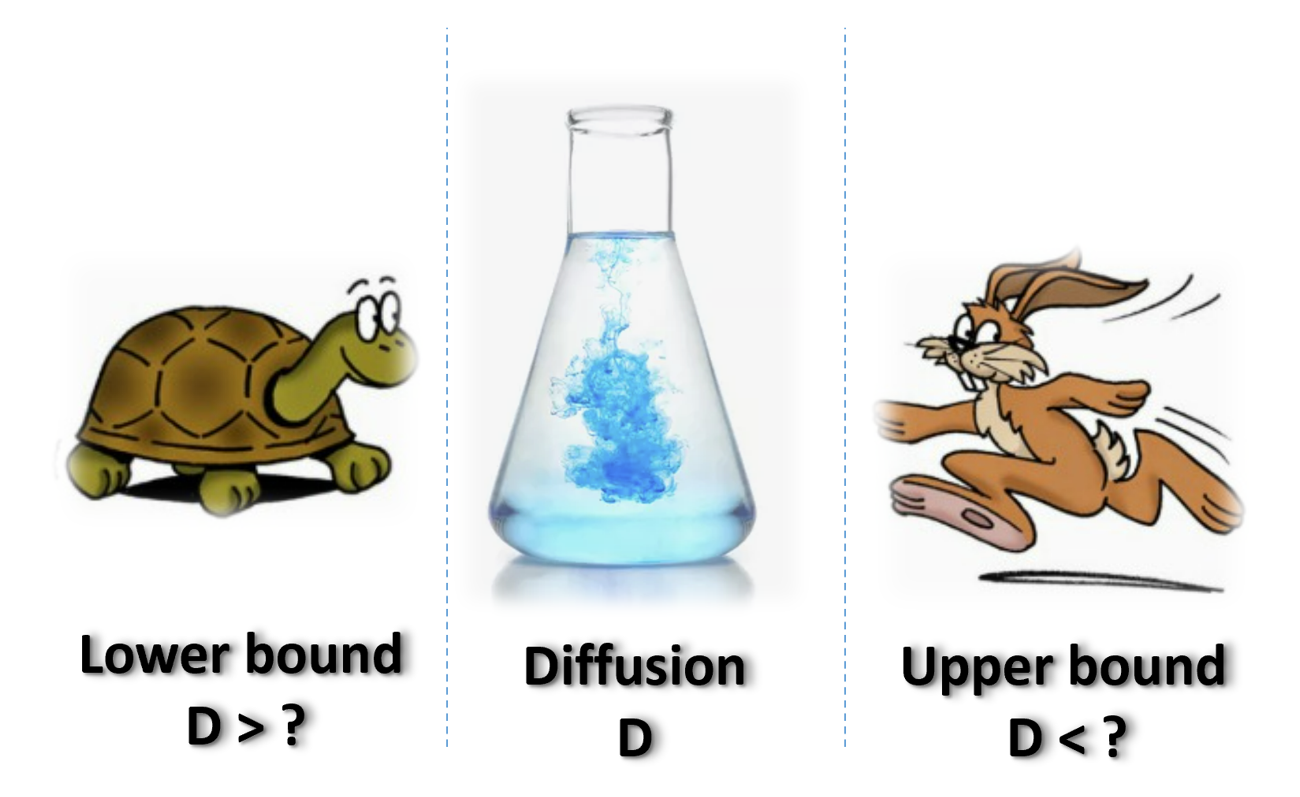}
     \caption{Given an arbitrary diffusive process and its related diffusion constant $D$, does $D$ obey any universal lower/upper bounds and in terms of which physical quantities? This is the question we address in this note.}
     \label{fig:0}
 \end{figure}\\
 Taking a similar perspective, one could ask whether the diffusion constants are also bounded from above or they can grow indefinitely (see cartoon in Fig.\ref{fig:0}). It turns out  that causality, and in particular the requirement of avoiding superluminal propagation, imposes a strong upper bound on diffusion which takes the form \cite{Hartman:2017hhp}:
 \begin{equation}
     D_i\,\leq\,v^2_{\text{lightcone}}\,\tau_{eq},\label{upperbound}
 \end{equation}
 where $v_{\text{lightcone}}$ is the velocity, setting the causal lightcone in the theory, and $\tau_{eq}$ the equilibration time at which the system thermalizes, and after which hydrodynamics starts to apply. The equilibration time can be universally defined using the imaginary part of the first damped, and therefore non-hydrodynamic, mode as:
 \begin{equation}
     \tau^{-1}_{eq}\,\equiv\,\omega_{eq}=|\mathrm{Im}\,\omega_{1}|,
 \end{equation}
 where $\omega_1$ is the frequency of the lowest of those modes.\\
 
 Contrary to the equilibration time, the definition of the lightcone velocity related to the causal structure is far from trivial in systems which do not enjoy relativistic invariance and/or systems with emergent IR lightcone structures. In relativistic systems, the lightcone velocity is obviously set by the speed of light $c$. One simple example is given by Israel-Stewart relativistic hydrodynamics \cite{ISRAEL1979341}. There, the lightcone speed is immediately identified with the speed of light $c$ and the equilibration time with the IR relaxation time $\tau_\pi$ which is pheomenologically introduced in the framework. Indeed, within the Israel-Stewart formalism, the absence of superluminality can be re-written exactly as an upper bound on the shear diffusion constant $D_{\text{shear}}<c^2 \tau_\pi$ \cite{Hartman:2017hhp,Baggioli:2020ljz}.\\
 
 A first check of the upper bound in Eq.\eqref{upperbound} was performed in Ref.\cite{Baggioli:2020ljz} using different velocity scales. A more formal derivation, based on technical mathematical properties of the hydrodynamic perturbative expansion, was discussed in \cite{Grozdanov:2020koi}. Interestingly,
 if one considers the momentum diffusion constant, instead of the $\eta/s$ ratio, all the known violations related to the breaking of spacetime symmetries disappear \cite{Baggioli:2020ljz}.\\
 
 Given the important role of hydrodynamics, recently, Ref.\cite{Arean:2020eus} proposed to connect the lightcone velocity with the equilibration velocity. In particular, Ref.\cite{Arean:2020eus} proposed a new bound which takes the following form:
 \begin{equation}
     D_i\,\leq\,v^2_{i,eq}\,\tau_{eq}\label{ublaise}\,.
 \end{equation}
 Here, $v_{i,eq}$ is the equilibration velocity for the $i^{th}$ diffusive mode, defined as:
 \begin{equation}
     v_{eq}\,\equiv\,\frac{\omega_{eq}}{k_{eq}}\,.
 \end{equation}
 where, to avoid clutter, the $i$ index has been dropped.\\
 This idea uses the recent definition of the radius of convergence of hydrodynamics presented in \cite{PhysRevLett.122.251601,Grozdanov2019,Withers:2018srf} and discussed further in \cite{Abbasi:2020ykq,Jansen:2020hfd,Baggioli:2020loj}. In particular, the pair $(\omega_{eq},k_{eq})$ corresponds to the location of the first (closest to the origin) critical point of the hydrodynamic perturbative series. This point coincides with the collision (in general in the complex plane) between a first (in this case diffusive) hydrodynamic mode and a nearby non-hydrodynamic mode (or a tower of them), and it determines the radius of convergence of the hydrodynamic series. More precisely, we utilize the following definitions:
 \begin{equation}
     k_{eq}\,\equiv\,|k^*|\,,\qquad \omega_{eq}\,\equiv\,|\omega^*|\,,
 \end{equation}
 where $(\omega^*,k^*)$ is the first critical point -- the position of the collision for complex frequency and momentum, $k^*,\omega^* \in \mathbb{C}$. In simple words, such a point determines the scale at which considering only conserved quantities is not enough anymore and the hydrodynamics description must be improved.\\
 
 Few comments are in order. (I) The definition of the equilibration velocity is specific to the diffusive process considered. In this sense, it is quite a stretch to consider the bound in Eq.\eqref{ublaise} as universal. Notice for example the crucial difference with the butterfly velocity proposal in Eq.\eqref{blakebound}, in which the velocity scale on the r.h.s. is the same for all the diffusion constants considered. (II) It is not clear how the bound in Eq.\eqref{ublaise} connects with that in Eq.\eqref{upperbound}. In particular, Eq.\eqref{ublaise} does not follow from the requirement of causality and furthermore $v_{eq}$ does not define nor the lightcone velocity nor the causal structure of any propagating process. (III) The equal sign in Eq.\eqref{ublaise} follows trivially from assuming that the diffusive dispersion relation:
 \begin{equation}
     \omega\,=\,-i\,D\,k^2
 \end{equation}
 is valid until the critical point $(\omega^*,k^*)$.\\
 In particular simple algebra gives
 \begin{align}
     \omega_{eq}\,=\,\,D\,k_{eq}^2\,\rightarrow\,1\,=\,\,D\,\frac{\omega_{eq}}{v_{eq}^2}\,\rightarrow\, D=v_{eq}^2 \tau_{eq}\,.
 \end{align}
 That said, the observation of \cite{Arean:2020eus} is interesting and it boils down to understand the following questions:
 \begin{itemize}
     \item Are there situations where the corrections to the hydrodynamic dispersion relation can be neglected until the critical point determining the radius of convergence of linearized hydrodynamics?
     \item Which conditions ensure the existence of such a scenarios and what is their meaning?
 \end{itemize}
 In this note, we consider the proposal of Ref.\cite{Arean:2020eus} in homogeneous holographic models with long-range order, i.e. with spontaneously broken translational invariance. These systems display an $\text{AdS}_2$ IR near-horizon geometry and a peculiar new diffusive mode labelled crystal diffusion \cite{Donos:2019txg,Baggioli:2020nay,Baggioli:2020haa}. In these holographic models, the lower bound Eq.(\ref{blakebound}) for crystal diffusion has already been verified in \cite{Baggioli:2020ljz}. Our task now is to determine whether the inequivalence in Eq.\eqref{ublaise} applies also to the same mode and which are the scales involved. Moreover, we analyze the connections and interplay between the lower bound on diffusion dictated by quantum chaos and this new upper bound determined by the breakdown of the hydrodynamic perturbative expansion. Finally, we provide some comments and thoughts for the future.

\section{The holographic model}
\begin{figure}[t]
    \centering
    \includegraphics[width=
   0.65 \linewidth]{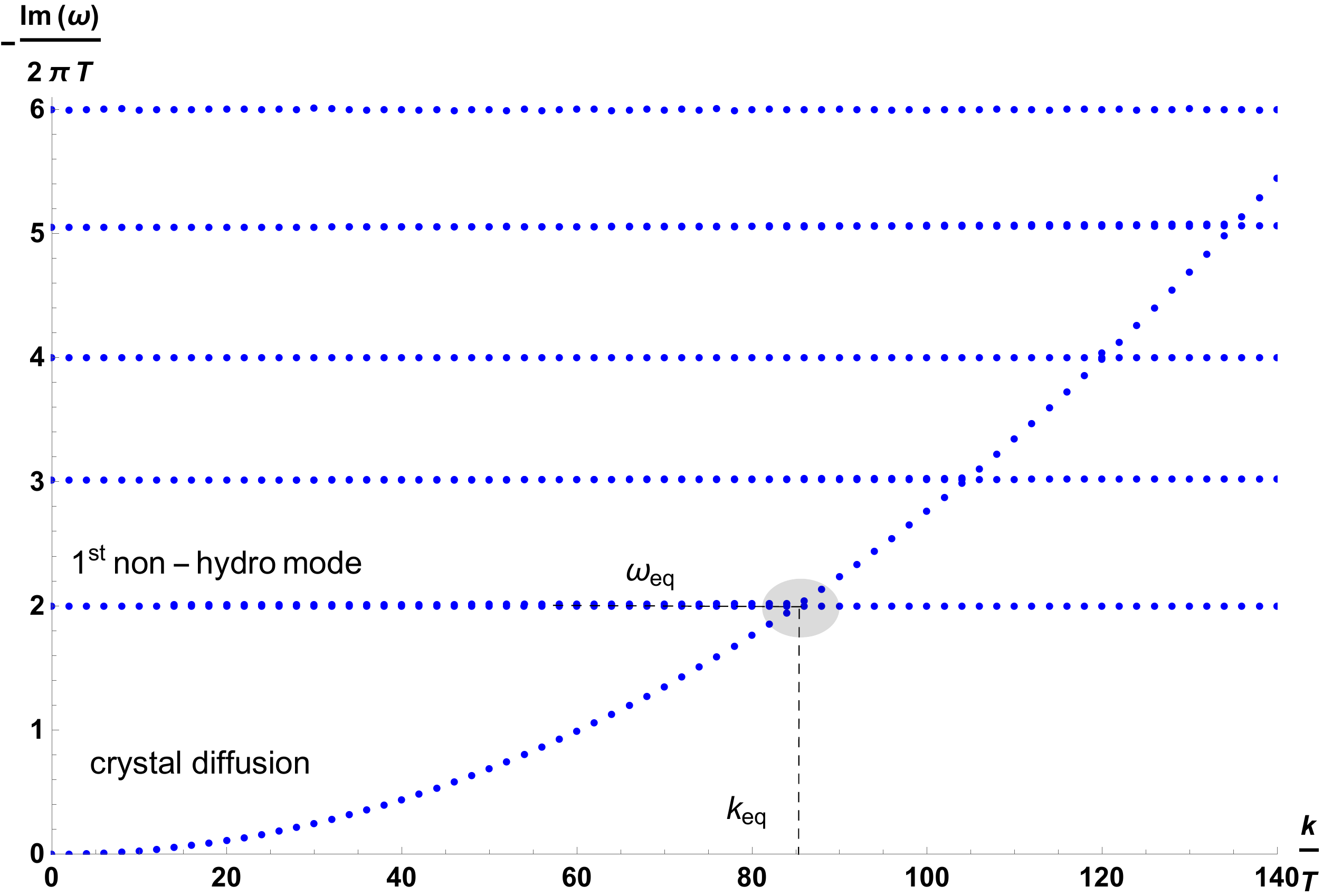}
    \caption{The longitudinal spectrum of quasinormal modes for $N=3$ and $m/T=500$. For simplicity, the attenuation constant of the longitudinal sound modes has been manually removed. The gray region emphasizes the location of the collision between the crystal diffusion mode and the first non-hydrodynamic $\text{AdS}_2$ mode, from which the parameters $\omega_{eq},k_{eq}$ are extracted as shown by the dashed lines. The tower of non-hydrodynamic modes follow the expected trend $\omega_n= 2\pi T (n+\Delta)$, where in this case $\Delta=2$. Similar pictures can be obtained for different values $N$ and  of $m/T$.}
    \label{fig:1}
\end{figure}
We consider a large class of holographic axion models \cite{Baggioli:2021xuv} introduced and discussed in~\cite{Baggioli:2014roa,Alberte:2015isw,Baggioli:2016rdj,Baggioli:2019rrs} and defined as follows:
\begin{equation}\label{action}
S\,=\,\int d^4x \sqrt{-g}
\left[\frac{R}2+ 3- \, m^2\,V(X)\right]\, ,
\end{equation}
where $X \equiv \frac12 \, g^{\mu\nu} \,\partial_\mu \phi^I \partial_\nu \phi^I$ and we have set the AdS radius to be unit. 
We choose an isotropic profile for the axion fields given by
\begin{equation}
    \phi^I\,=\,x^I\,,
\end{equation}
which represents a trivial solution of the equations of motion because of the global shift symmetry $\phi^I\rightarrow \phi^I+b^I$ of the action \eqref{action}.
The background geometry in Eddington-Filkenstein coordinates is written as:
\begin{equation}
\label{backg}
ds^2=\frac{1}{u^2} \left[-f(u)\,dt^2-2\,dt\,du + dx^2+dy^2\right]\, ,
\end{equation}
where $u\in [0,u_h]$ is the radial holographic direction going from the boundary $u=0$ to the horizon, $f(u_h)=0$.
Finally, we have:
\begin{equation}\label{backf}
f(u)= u^3 \int_u^{u_h} dv\;\left[ \frac{3}{v^4} -\frac{m^2\,V(v^2)}{v^4}\right] \, ,
\end{equation}
and consequently the temperature $T$ is defined as
\begin{equation}
T=-\frac{f'(u_h)}{4\pi}=\frac{6 -  2 m^2\, V\left(u_h^2 \right)}{8 \pi\, u_h}\, ,\label{eq:temperature}
\end{equation}
while the entropy density is given by $s=2\pi/u_h^2$.\\

In the rest of the note, we will focus on the monomial form:
\begin{equation}
    V(X)\,=\,X^N\label{bench}\,.
\end{equation}
We refer to the previous literature \cite{Alberte:2017cch,Alberte:2017oqx,Baggioli:2018vfc,Andrade:2019zey,Ammon:2019wci,Ammon:2019apj,Baggioli:2019abx,Ammon:2020xyv} for more details concerning these models and their properties.\\
In these holographic models, the UV expansion of the $\phi^I$ bulk fields reads:
\begin{equation}
    \phi^I(t,x,u)\,=\, \phi_{(0)}^I(t,x)\,+\,\phi_{(1)}^I(t,x)\,u^{5-2N}\,+\,\dots
\end{equation}
Assuming standard quantization, one could verifiy that for $N<5/2$ the background solution $\phi^I=x^I$ plays the role of an external source, while for $N>5/2$ it describes a finite expectation value for the operators $\mathcal{O}^I$, dual to the axion fields. Following this argument, the breaking of translations is explicit for $N<5/2$ and spontaneous for $N>5/2$. In this note, we will only consider the case $N>5/2$.
\section{Results}
The longitudinal spectrum of systems with spontaneously broken translations features a peculiar mode with diffusive dispersion relation which is usually labelled ``crystal diffusion". Using the correct hydrodynamic description \cite{Armas:2019sbe}, the diffusion constant of such a mode reads:
\begin{equation}
D\,=\,\xi\,
  \frac{\left(B+G-\mathcal{P}\right)\,\chi_{\pi\pi}}{s'\,T^2\,v_L^2}\,.\label{hform}
\end{equation}
The various parameters appearing in the equation above are: the Goldstones diffusion constant $\xi$, the bulk modulus $B$, the shear modulus $G$, the momentum susceptibility $\chi_{\pi\pi}$, the temperature derivative of the entropy $s'$ and finally the so called crystal pressure $\mathcal{P}$. The hydrodynamic formula Eq.\eqref{hform} has been successfully matched to the holographic results \cite{Ammon:2020xyv,Baggioli:2019abx} after some initial, and then resolved, tension \cite{Ammon:2019apj}.\\

\begin{figure}
    \centering
    \includegraphics[width=
  0.6  \linewidth]{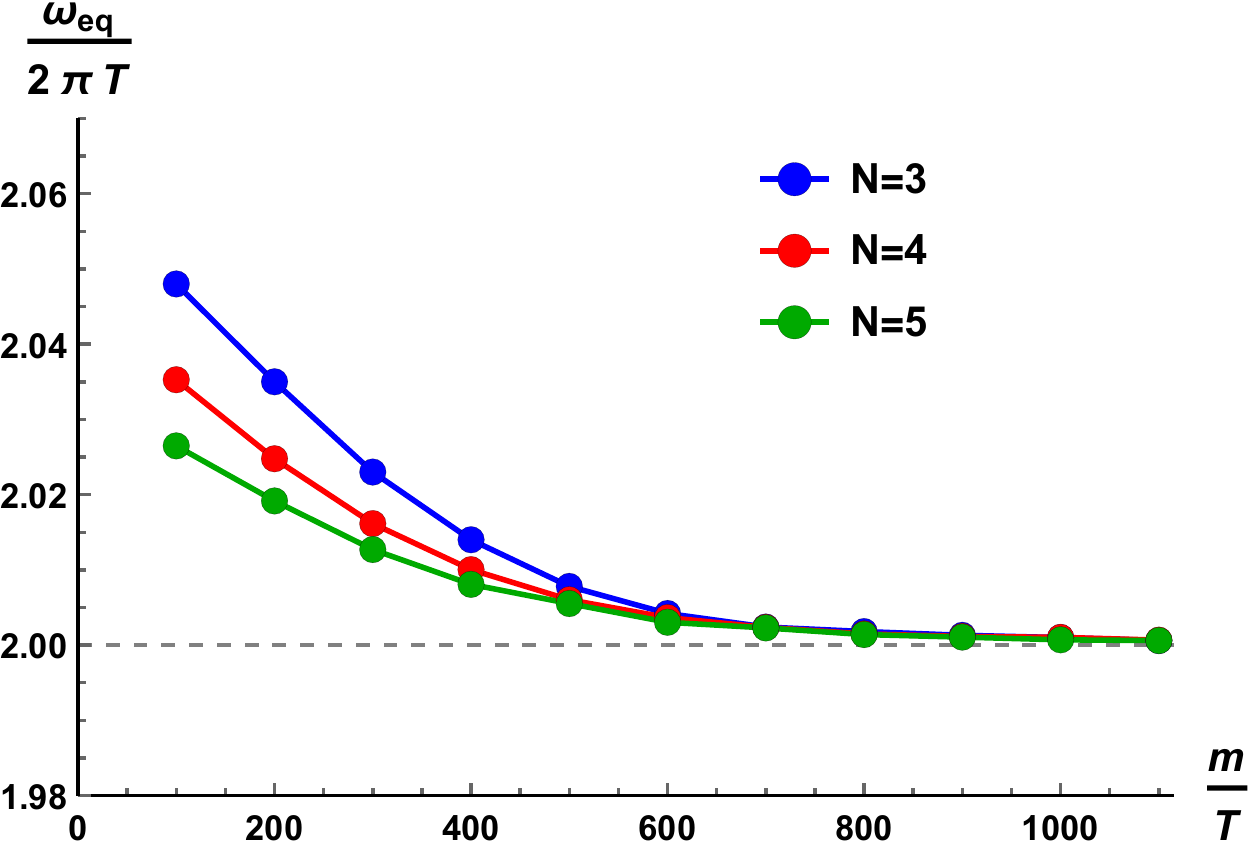}
    \caption{The equilibration frequency $\omega_{eq}$ in function of the dimensionless temperature for various $N=3,4,5$. All the curves approach asymptotically a constant value given by $\Delta_0=2$.}
    \label{fig:2}
\end{figure}

The dispersion relation of the crystal diffusion mode is shown in Fig.\ref{fig:1} for very low values of temperature. The characteristic quadratic scaling and the corresponding slope are consistent with the previous theoretical computations. In Fig.\ref{fig:1}, the lowest non-hydrodynamic modes $\omega=-i\, \omega_n$ are shown as well. As explained in the previous literature \cite{Faulkner:2009wj,Edalati:2010hk} and discussed in \cite{Arean:2020eus}, these modes have a quite flat dispersion relation at low temperatures and they appear equally separated. In particular, this tower of non-hydrodynamic modes belongs to the IR $\text{AdS}_2$ spectrum and it is given by:
\begin{equation}
    \omega_n\,=\,2\,\pi\,T\,\left(\Delta_0\,+\,n\right)\qquad \text{for}\qquad T\rightarrow 0\,,\label{dim}
\end{equation}
where $n$ is the index labelling the modes and the constant $\Delta_0$ corresponds to the conformal dimension of the lowest operator in the $\text{AdS}_2$ IR fixed point evaluated at zero momentum, $\Delta_0=\Delta(k=0)$\cite{Arean:2020eus}.\\
Given the presence of these modes, we can immediately identify the value of the lowest ($n=0$) frequency in Eq.\eqref{dim} with the equilibration timescale:
\begin{equation}
    \omega_{eq}=\tau^{-1}_{eq}=\omega_0=2\pi\,\Delta_0\,T\,.
\end{equation}
We plot the value of the normalized equilibration frequency in function of $m/T$ at low temperatures in Fig.\ref{fig:2}. The curves approach nicely a common asymptotic $T=0$ value which is given by $\Delta_0=2$. Importantly, this value is completely independent of the choice of the potential and in particular the value of the power $N$ \footnote{We are grateful to Hyun-Sik Jeong, Keun-Young Kim and Ya-Wen Sun for pointing out a mistake in a previous version of our manuscript which was due to numerical inaccuracies. For an analytic derivation, and extension, of our results we refer to their forthcoming work.}.
\\

\begin{figure}[ht]
    \centering
    \includegraphics[width=
    \linewidth]{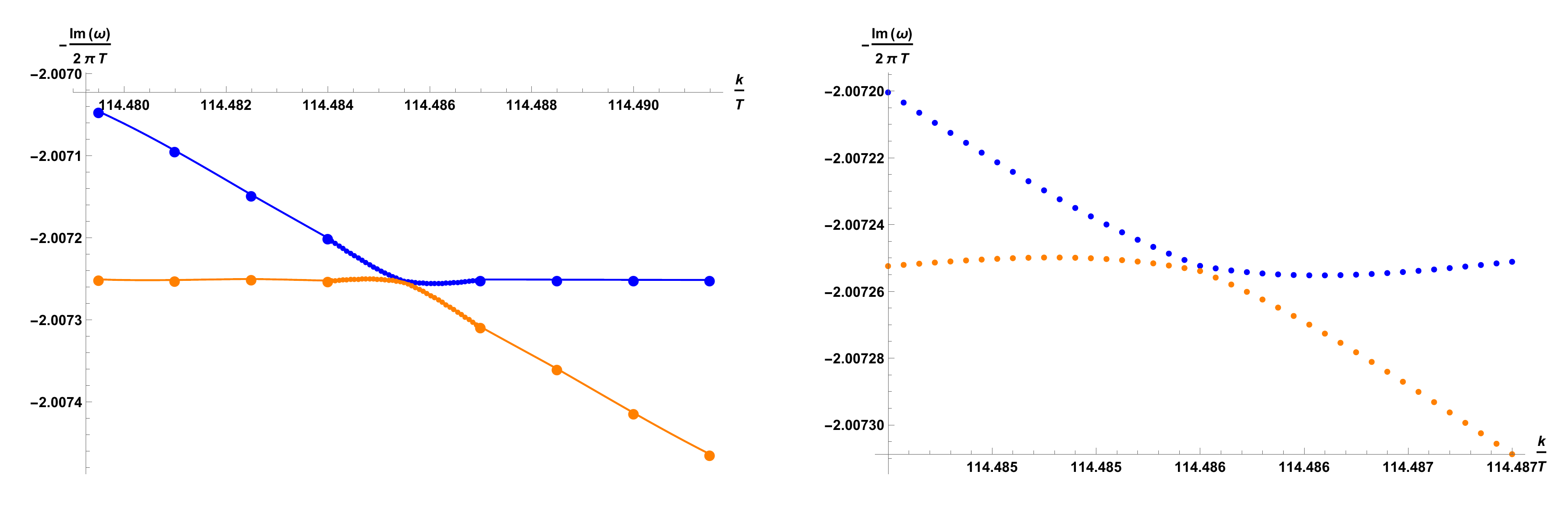}
    
    \vspace{0.2cm}
    
    \includegraphics[width=0.45 \linewidth]{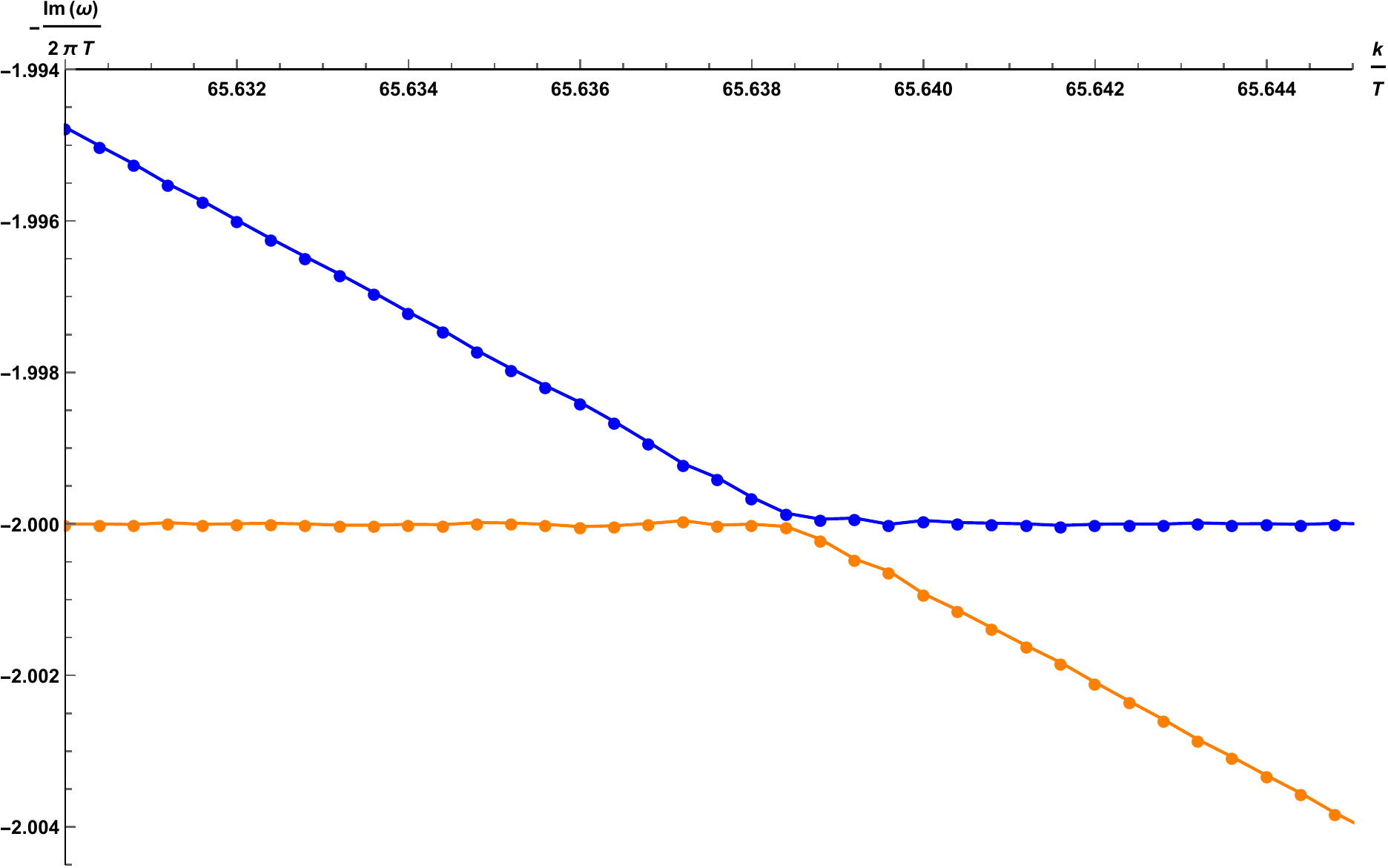}
    \qquad 
     \includegraphics[width=0.45 \linewidth]{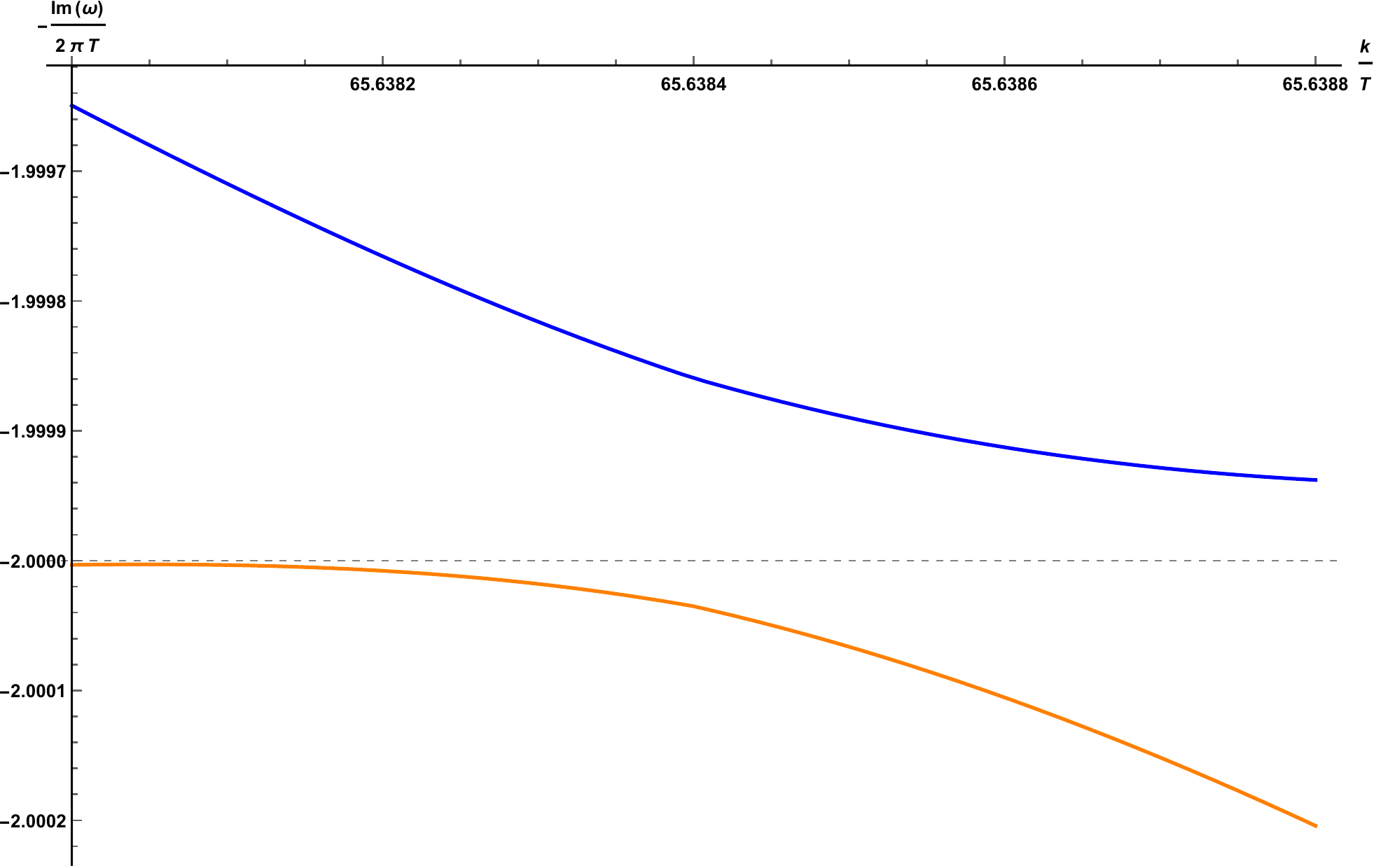}
    \caption{\textbf{Top Left: } The "almost" collision between the diffusive mode and the first non-hydrodynamic mode. \textbf{Top Right: }A zoom of the avoided crossing dynamics for $N=3$ and $m/T=900$. The gap between the two modes becomes smaller and smaller going to lower temperature. \textbf{Bottom panels: } The same figures at larger temperature, $m/T=300$. Here, the avoided crossing is more pronounced and the separation between the two curves is larger. This implies that the imaginary part of the critical complex momentum is larger.}
    \label{fig:zoom}
\end{figure}

Let us now move to discuss the interplay between the diffusive hydrodynamic mode and the first non-hydro mode. We have zoomed in the area in which the crystal diffusion mode and the first non-hydrodynamic mode approach each other, which is indicated with a gray shaded region in Fig.\ref{fig:1}. The results are more clearly shown in Fig.\ref{fig:zoom}. From there, it is evident that the two modes display an avoided crossing dynamics as noticed already in \cite{Arean:2020eus}. In particular, we numerically observe (see \cite{Arean:2020eus} for an analytic proof) that the avoided crossing mechanism becomes more and more evident by increasing the temperature. This is consistent with the observation that the collision between the diffusive mode and the first non-hydrodynamic mode happens for complex momentum, but the imaginary part of the critical point tends to zero with the temperature $T$. Therefore, at low temperature, we can approximate the equilibration scale with the real value of the critical momentum:
\begin{equation}
    k_{eq}\,\equiv\,|k^*|\,\approx \mathrm{Re}\,k^*
\end{equation}
which, together with the critical frequency $\omega^*$, is easily readable from the dispersion relation of the modes in the longitudinal spectrum, as shown in  Fig.\ref{fig:1}.
\\

\begin{figure}
    \centering
    \includegraphics[width=
  0.45  \linewidth]{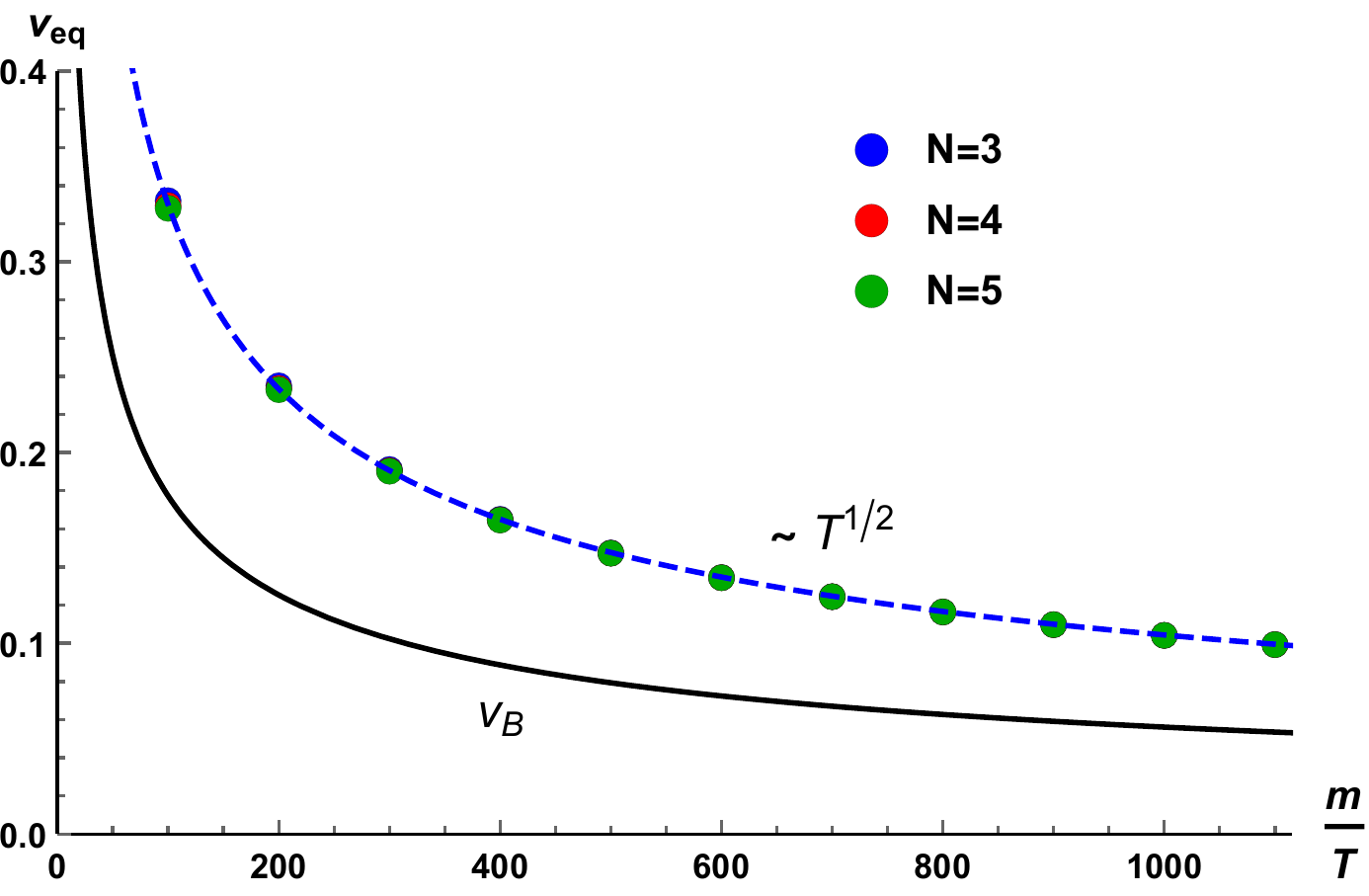}
    \qquad
    \includegraphics[width=
    0.45\linewidth]{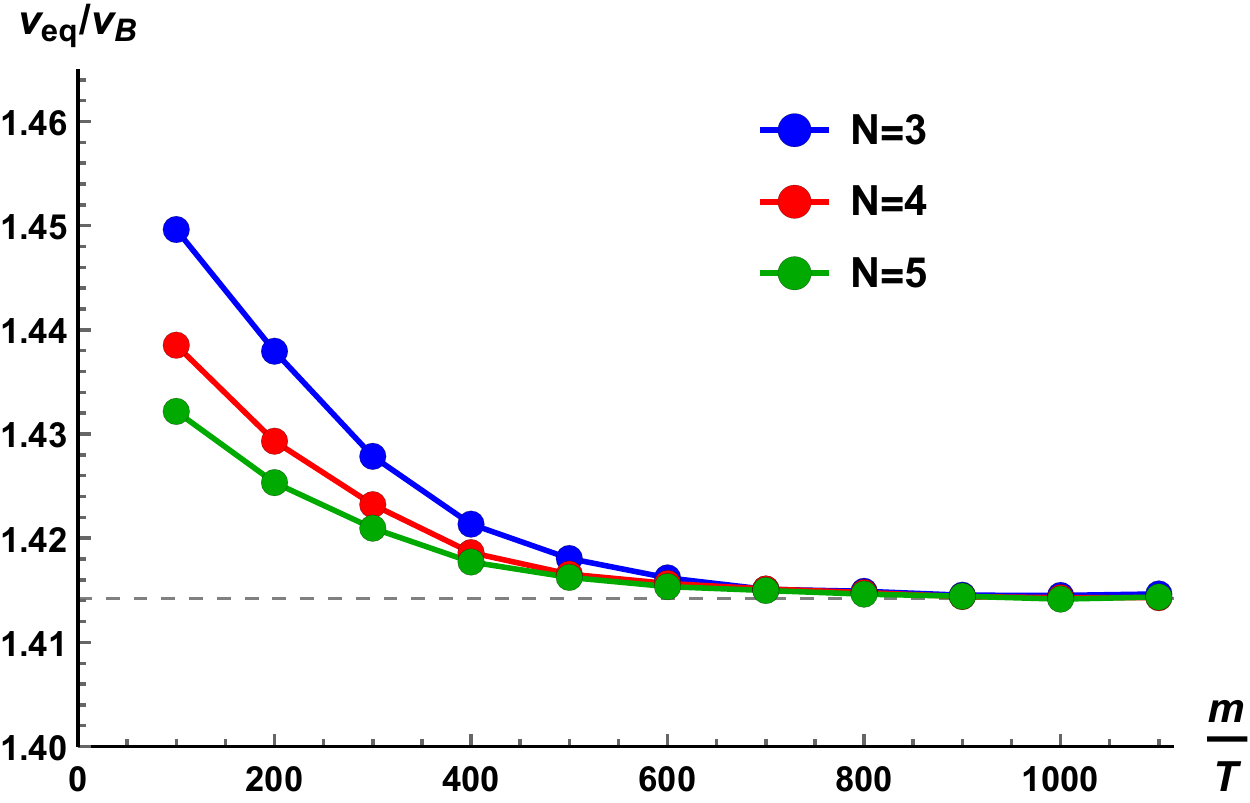}
    \caption{\textbf{Left: }The behaviour of the equilibration velocity in function of the dimensionless temperature $m/T$ for various values of $N=3,4,5$. The dashed line shows the $T^{1/2}$ scaling. \textbf{Right: } The comparison with the butterfly velocity. At low temperatures, the ratio of the two velocities approach a constant value, $\sqrt{\Delta_0}=\sqrt{2}$.}
    \label{fig:3}
\end{figure}

In summary, we can extract both parameters, $\omega_{eq},k_{eq}$, simply by looking at the dispersion relations of the two lowest modes as shown in Figures \ref{fig:1} and \ref{fig:zoom}. At this point, we can also straightforwardly define the equilibration velocity as:
\begin{equation}
    v_{eq}\,\equiv\,\frac{\omega_{eq}}{k_{eq}}\,.\label{vdef}
\end{equation}
We show the behaviour of the non-normalized equilibration velocity in function of temperature in the left panel of Fig.\ref{fig:3}. Interestingly, we observed a very clear $T^{1/2}$ scaling close to zero temperature. Given that $\omega_{eq}\sim T$ at low temperature (Eq.\eqref{dim}), we can derive that in such a regime the radius of convergence of hydrodynamics goes as:
\begin{equation}
    k_{eq}\,\sim\,T^{1/2}\,. \label{find}
\end{equation}
This is consistent with the idea that the convergence properties of hydrodynamics become worse and worse upon lowering the temperature \cite{Baggioli:2020loj}. In other words, hydrodynamics breaks down at larger distances going towards zero temperature. Eq.\eqref{find} is also consistent with previous results for charged holographic backgrounds and realistic liquids \cite{Abbasi:2020ykq,Jansen:2020hfd,Baggioli:2020loj}.
\\

At this stage, we want to compare the behaviour of the equilibration velocity with that of the butterfly velocity, which has a fundamental role in the diffusion bounds discussed in the introduction. The butterfly velocity can be obtained from horizon data and in our background is given by:
\begin{equation}
    v_B^2\,=\,\frac{\pi\,T}{u_h}\,.
\end{equation}
In the limit of $m/T \rightarrow \infty$, the radius of the horizon goes to a constant and therefore we obtain the expected scaling $v_B\sim T^{1/2}$. Interestingly, the ratio between the two velocity scales approaches a constant, given by $\sqrt{\Delta_0}$, in the low temperature limit. We notice that for canonically normalized operators with unitary conformal dimension in the $AdS_2$ IR fixed point, the two velocities would exactly coincide at low temperature. More in general, we find that:
\begin{equation}
    v_{eq}\,>\,v_B\,.
\end{equation}
If one considers the butterfly velocity as an emergent lightcone speed of some (non-relativistic) quantum chaotic system, the relation just obtained looks quite dangerous since it would imply a superluminal propagation with speed $v_{eq}$ outside of the causal lightcone. Nevertheless, the equilibration velocity $v_{eq}$ does not correspond to any propagating modes. In other words, in these cases, there is absolutely no excitation propagating at such speed. In any case, this relation looks certainly interesting and it deserves further understanding. It also implies that the thermalization speed, defined as in Eq.\eqref{vdef}, is faster than the speed of information scrambling. It would be important to understand how universal this hierarchy is and which are the physical consequences.\\

Another indication that the equilibration speed $v_{eq}$ cannot play the role of the lightcone velocity in the bound of \cite{Hartman:2017hhp} is given by the fact that in this model the longitudinal speed of sound is much larger than the equilibration velocity. In this sense, if one had to choose a lightcone velocity, the sound speed would be the most appropriate.\\

\begin{figure}[ht]
    \centering
    \includegraphics[width=
   0.6\linewidth]{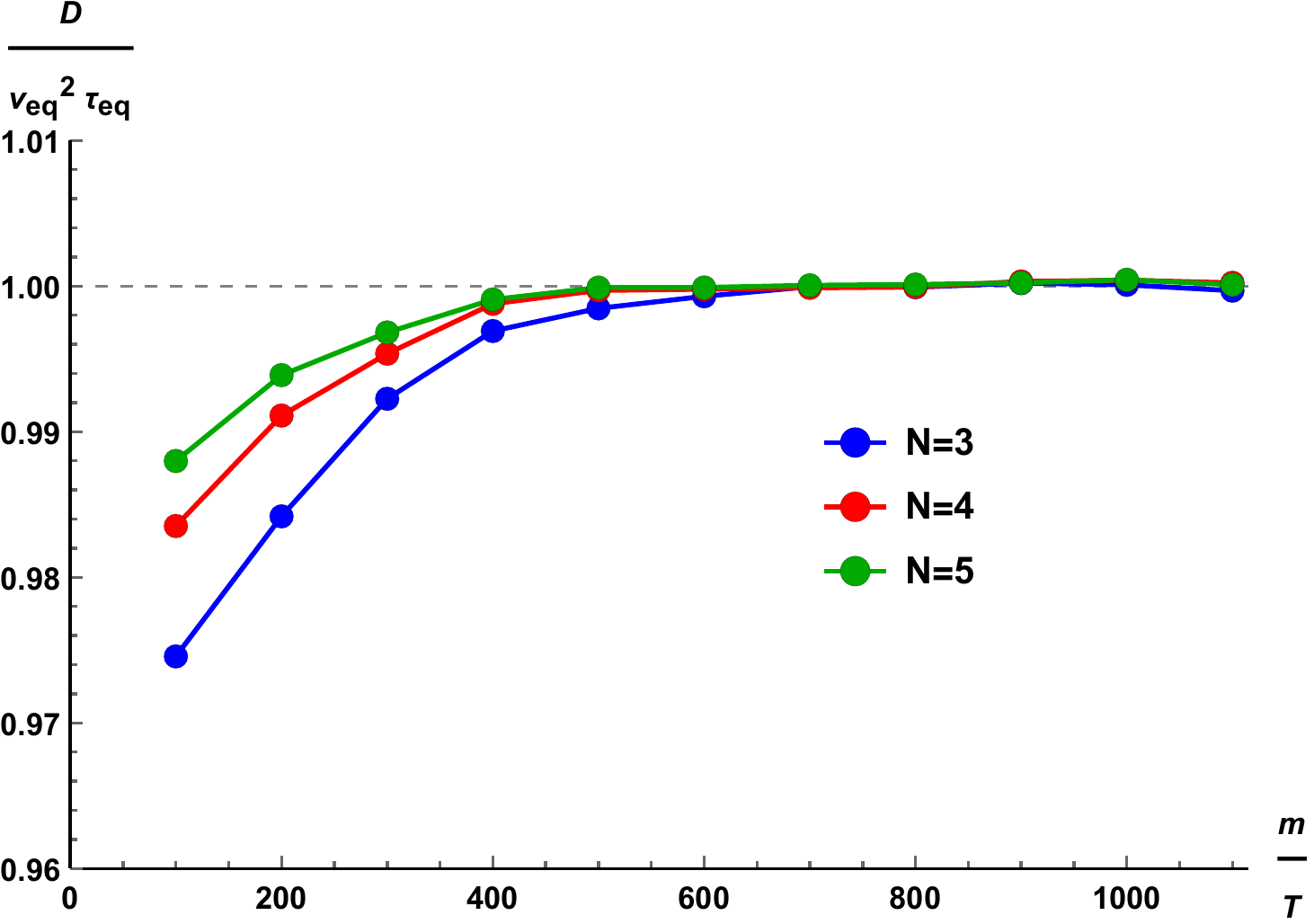}
    \caption{The dimensionless ratio $D/(v_{eq}^2\tau_{eq})$ in function of the inverse dimensionless temperature for various powers $N=3,4,5$. All the curves approach the unit value at low temperature. Moreover, all curves are consistent with the inequality $D\leq v_{eq}^2\tau_{eq}$.}
    \label{fig:4}
\end{figure}

After having identified all the scales entering in the bound of Eq.\eqref{ublaise}, we can finally test its validity for the crystal diffusion mode. In Fig.\ref{fig:4}, we plot the dimensionless ratio $D/v_{eq}^2\tau_{eq}$ in function of the dimensionless inverse temperature $m/T$ at small temperatures and for various potentials, $N=3,4,5$. In all the cases, we notice that this ratio approaches unity at $T\rightarrow 0$ confirming the validity of the relationship:
\begin{equation}
    D\,=\,v_{eq}^2 \tau_{eq}\qquad \text{for}\qquad T \rightarrow 0\,.
\end{equation}
Moreover, we obtain that in general 
\begin{equation}
    D\,\leq\,v_{eq}^2 \tau_{eq}
\end{equation}
and that therefore the bound Eq.\eqref{ublaise} proposed in \cite{Arean:2020eus} indeed holds.
This constitutes an explicit confirmation that even for the crystal diffusion mode, the diffusion constant is bounded from above by the equilibration scales $v_{eq}$ and $\tau_{eq}$.
\section{Conclusions}
In summary, in this short note, we have confirmed the validity of the diffusion bound defined in terms of the hydrodynamics breakdown data proposed in Ref.\cite{Arean:2020eus} for the crystal diffusion mode present in all holographic systems with spontaneously broken translations. Importantly, in our case, the collision determining the equilibration scales happens for real values of the momentum $k$ and therefore it is easily extracted from the dispersion relation of the lowest quasinormal modes.\\

\begin{figure}
    \centering
    \includegraphics[width=
   0.6 \linewidth]{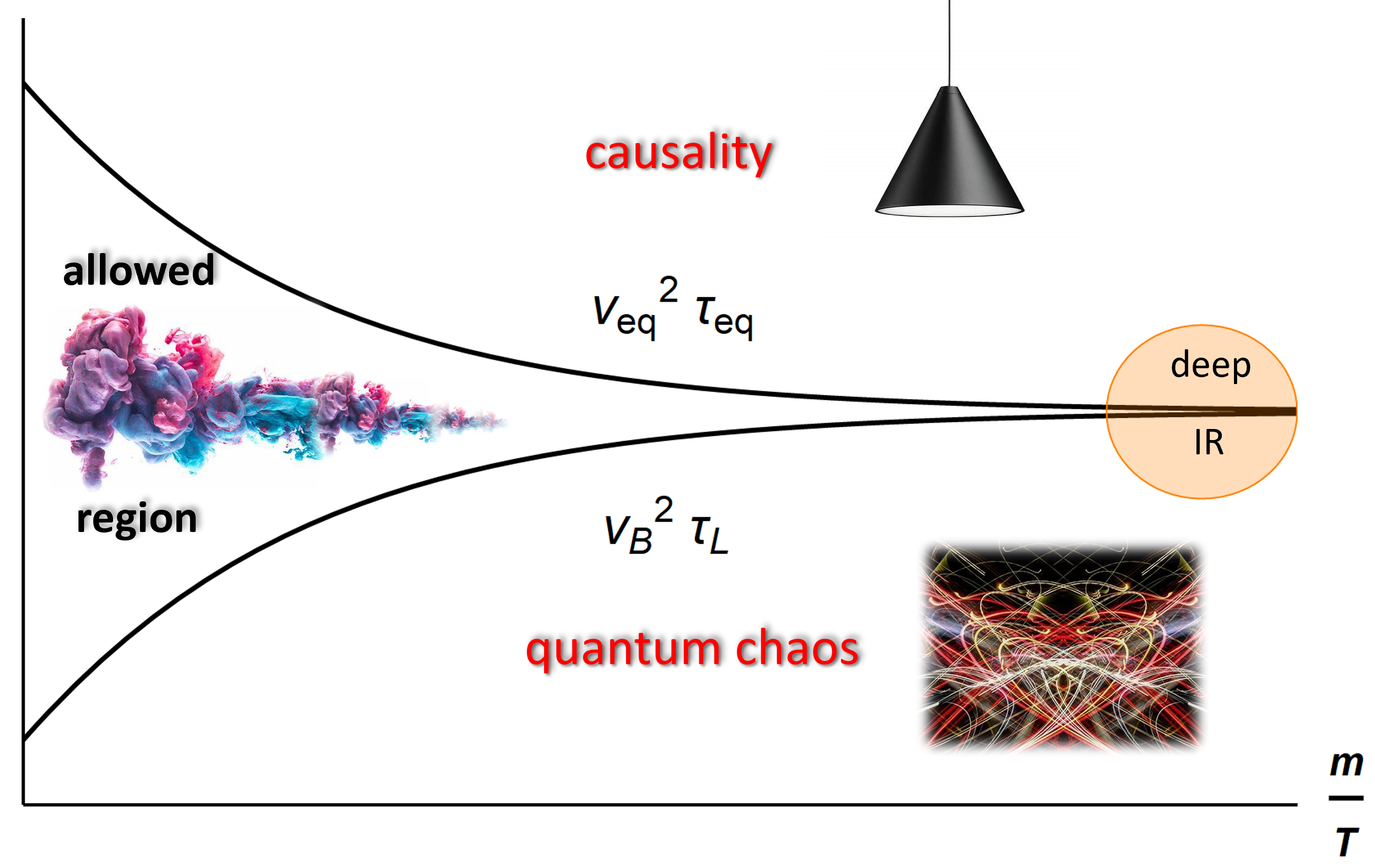}
    \caption{A cartoon of the allowed region for the diffusion constant $D$. The upper edge comes from the scales related to the breakdown of linearized hydrodynamics. While, the lower one is dictated by the data of quantum chaos \cite{Baggioli:2020ljz}. Both curves merge in the limit of $T\rightarrow 0$.}
    \label{fig:5}
\end{figure}

Moreover, we do find that, for arbitrary values of the temperature, the diffusion constant is confined in an range determined by:
    \begin{equation}
        v_B^2\,\tau_L\,\leq\,D\,\leq\,v_{eq}^2\,\tau_{eq},
    \end{equation}
    which is shown in Fig.\ref{fig:5}. Importantly, approaching the zero temperature limit this allowed region shrinks and at exactly zero temperature the two limits collapse on each other. This means that the universal bounds determine uniquely the value of the diffusion constant at zero temperature:
    \begin{equation}
       \left( v_B^2\,\tau_L\right)|_{T=0}\,=\,D|_{T=0}\,=\,\left(v_{eq}^2\,\tau_{eq}\right)|_{T=0}\,.\label{zeroT}
    \end{equation}
    This is a very interesting and new outcome whose universal character must be investigated further.
\section{Additional comments}
We conclude with several comments and ideas for the future.
\begin{itemize}
    \item One interesting point is the connection between the diffusivity bound $D\leq v_{eq}^2\tau_{eq}$ and the higher order corrections to the diffusive dispersion relation. As we have shown in the main text, if the diffusive behaviour $\omega=-i D k^2$ persists until the collision point, the equality trivially holds. What happens if that is not the case? In general, the dispersion relation is given by a perturbative expansion in momentum of the type:
    \begin{equation}
        \omega\,=\,-i 
        \left(D k^2+ a_1\,k^4\,+\,a_2\, k^6+\dots\,\right),
    \end{equation}
    where for simplicity we have ignored any possible real part and considered a purely imaginary mode. Let us consider that, upon reaching the critical point $(\omega^*,k^*)$, the first higher order correction cannot be neglected. Then, we have:
    \begin{equation}
        \omega_{eq}\,=\,
       D k_{eq}^2+ a_1\,k_{eq}^4
    \end{equation}
    which, after some manipulations, gives:
    \begin{equation}
        D\,=\,v_{eq}^2\tau_{eq}\,-\,\frac{a_1}{v_{eq}^2\,\tau_{eq}}\,.\label{look}
    \end{equation}
    The extension to higher order terms is trivial and for simplicity not shown here.
   By looking at Eq.\eqref{look}, one can infer that the validity of the upper diffusion bound depends crucially on the signs of the higher order corrections. In particular, the higher order terms in the diffusive dispersion relations introduce corrections of order $\mathcal{O}(1/v_{eq}^{2n} \tau_{eq}^n)$, controlled by the various new coefficients $a_n$. Is there any physical requirement (e.g. causality) that fixes the sign of these coefficients and therefore the validity of the upper bound? That seems indeed the case. The univalence property of the hydrodynamics expansion put stringent bounds on all the higher order coefficients \cite{Grozdanov:2020koi}. For example, it constraints the first of them, labelled $a_1$ above, to be positive. In other words, it is very tempting to claim that the validity of the upper bound proposed in \cite{Arean:2020eus} can be formally derived using mathematical properties of the hydrodynamic series \cite{Grozdanov:2020koi}. A simple scenario where this mechanism appears is the telegrapher equation \cite{BAGGIOLI2020}:
   \begin{equation}
       \omega^2+i \omega/\tau\,=\,v^2\,k^2\,.\label{tel}
   \end{equation}
   In this case, the purely diffusive dispersion law gets corrected before the poles collision as:
   \begin{equation}
       \omega\,=\,-\,i\,D\,k^2\,-\,i\,v^4\,\tau^3\,k^4\,+\,\dots
   \end{equation}
   and the corresponding higher order coefficient reads $a_1=v^4\,\tau^3$. Stability, and more precisely the requirement of having $\tau>0$ (a relaxation process and not an ``exploding" one), implies that $a_1>0$ and therefore that the diffusion constant is bounded from above, as discussed in the previous paragraph.
    \item 
    It would be interesting to understand better the zero temperature relation in Eq.\eqref{zeroT}. To the best of our knowledge, the collapse of the two bounds at zero temperature has not been observed nor discussed before. How universal is this feature? What can we learn from it? Is it possible to maintain a finite range of allowed values at zero temperature or the two bounds always collapse?
    \item In all this discussion, the value of the constant $\Delta_0$ plays a fundamental role. In particular, the concrete value $\sqrt{\Delta_0}$ controls the ratio between the equilibration velocity and the butterfly velocity at low temperature. It would be interesting to understand if any physical requirement (e.g. unitarity and the corresponding bounds on the conformal dimensions) constraints $\Delta_0>1$ and if not what is the meaning of this role-reversal phenomenon.
    \item A priori, it is not clear what is the relation between the equilibration velocity $v_{eq}$ and the causal structure of the system. In particular, such a velocity in general does not correspond to any propagating mode. Interestingly, taking the telegrapher equation \eqref{tel}, one can derive that the equilibration velocity $v_{eq}$ coincides exactly with the sound speed of the emergent propagating mode at large momentum. In this simplified scenario, it does corresponding to a propagating mode at short distance. 
    \item In the main text, we have derived that in the low temperature limit, $\tau_{eq}^{-1}=2\pi\, \Delta_0\,T$. This relaxation time has the same temperature dependence of the Planckian time and the Liapunov time but with a different numerical prefactor. In particular we have:
    \begin{equation}
        (\tau_{pl},\tau_{L},\tau_{eq})\,T\,=\,\left(1,\frac{1}{2\pi},\frac{1}{2 \pi\,\Delta_0}\right)\,.
    \end{equation}
    Moreover, the hierarchy of these timescales depends crucially on the value of $\Delta_0$. Also, the situation might be substantially different away from maximal chaos \cite{Choi:2020tdj} where the Maldacena bound \cite{Maldacena:2015waa} is not saturated and $\tau_{L} > \frac{1}{2\pi\,T}$. Which is the order of these timescales and how can it be changed?
    \item In \cite{Arean:2020eus}, the equilibration velocity has been defined using the critical point which determines the breakdown of the hydrodynamics expansion and in particular of the diffusive dispersion relation. In principle, there are other points in the complex plane which assume a particular role -- the pole-skipping points \cite{Grozdanov:2017ajz,Blake:2018leo,Blake:2017ris,Ahn:2020baf}. Given a certain Green function, those are the points at which the zeros and the poles cross, rendering the Green function indeterminate. Following \cite{Arean:2020eus}, one could define a pole-skipping velocity:
    \begin{equation}
        v^{\mathcal{O}}_{skip}\,\equiv\,\frac{|\omega_{skip}|}{|k_{skip}|},
    \end{equation}
    where $(\omega_{skip},k_{skip})$ indicates the location of the first pole skipping point in the complex plane and the index $\mathcal{O}$ the operator whose Green function is considered. Notice that if one considers the energy-energy correlator, one  finds $v^{\epsilon}_{skip}=v_B$ and $\omega_{skip}=\tau_L^{-1}$ \cite{Grozdanov:2017ajz}. In this sense, for the energy correlator, we have the following identification:
    \begin{equation}
        \frac{v^{2}_{skip}}{\omega_{skip}}\,=\,v_B^2\,\tau_L
    \end{equation}
    and therefore one can immediately write down a lower bound of the type:
    \begin{equation}
        D_\epsilon\,\geq\,\frac{v^{2}_{skip}}{\omega_{skip}}\,.
    \end{equation}
    The question whether a more general bound:
    \begin{equation}
        D_{\mathcal{O}}\,\geq\,\frac{v^{2}_{\mathcal{O},skip}}{\omega_{\mathcal{O},skip}}\,
    \end{equation}
    exists for an arbitrary operator $\mathcal{O}$ is valuable and it can be easily investigated with the existing techniques. Preliminary indications \cite{private} seem to suggest our hypothesis.
    \item As already mentioned, the speed of longitudinal sound $v_L$ in this model is much larger than the equilibration speed $v_{eq}$. This implies that (I) the equilibration velocity cannot be taken as the one determining the causal lightcone and  (II) that the bound $D \leq v_{eq}^2 \tau_{eq}$ is more stringent than the one coming from causality as in \cite{Hartman:2017hhp}.
    \item Finally, it would be interesting to consider IR fixed point with dangerously irrelevant deformations \cite{Davison:2018ofp}. There, the equilibration time is expected to be parametrically longer, $\tau_{eq}\gg T^{-1}$, and the full picture could change substantially. Hyperscaling-Lifshitz IR geometries are also a straightforward generalization of this program.
 \end{itemize}
The emerging global picture suggests intriguing and possibly fundamental connections between transport, quantum chaos, hydrodynamics and pole skipping which are left to be revealed.\\

We plan to come back to some of these questions in the near future.
\subsection*{Acknowledgments} 
We thank Sebastian Grieninger for providing the numerical codes used in previous works and for useful comments. We thank Saso Grozdanov, Keun-Young Kim, Yongjun Ahn and Hyun-Sik Jeong for reading a preliminary version of the manuscript and providing useful comments and suggestions. We thank Keun-Young Kim, Yongjun Ahn, Hyun-Sik Jeong and Ya-Wen Sun for sharing with us unpublished results and for correcting a mistake contained in the previous version of this manuscript.
N.W. and W.J.L. are supported by NSFC No.11905024 and No.DUT19LK20.
M.B. acknowledges the support of the  Shanghai Municipal Science and Technology Major Project (Grant No.2019SHZDZX01).

\bibliography{diff}
\bibliographystyle{JHEP}

\appendix 
\newpage
\section{Equations for the perturbations}
We align the momentum $k$ along the $y$ direction. The perturbations in the longitudinal sector are given by
\begin{equation}
    \{h_{x,\,{s}}=1/2\,(h_{xx}+h_{yy}),\,h_{x,\,{a}}=1/2\,(h_{xx}-h_{yy}),\,\delta\phi_y,\, h_{tt},\, h_{ty}\}\,,
\end{equation}
We use a radial gauge. The final set of equations for the perturbations reads
\begin{align}
&u f'\, \delta \phi_y'\,  \dot{V}\,+2\, u^2\, f\, \delta \phi_y'\,  \ddot{V}+u\, f\, \delta \phi_y''\,  \dot{V}-2\, f\, \delta \phi_y'\,  \dot{V}-k^2\, u\, \delta \phi_y\,  \dot{V}-k^2\, u^3\, \delta \phi_y\, \ddot{V}\nonumber\\
&+i\, k\, u\, h_{x,\,{a}}\,  \dot{V}-i\, k\, u^3\, h_{x,\,{s}}\, \ddot{V}+2\, i\, u^2\, \omega\,  \delta \phi_y\, \ddot{V}+u\, h_{ty}'\,  \dot{V}+2\, i\, u\, \omega\,  \delta \phi_y'\, \dot{V}\nonumber\\
&-2\, i\, \omega\,  \delta \phi_y  \dot{V}-2\, h_{ty} \left(\dot{V}-u^2\, \ddot{V}\right)=0\\
&u (f \left(u f'\, h_{x,\,{s}}'-2\, f\, h_{x,\,{s}}'-2\, i\, k\, m^2\, u^3\, \delta \phi_y\, \ddot{V}-u\, h_{tt}''+4\, h_{tt}'\right)\nonumber\\
&+k\, h_{ty} \left(i\, u\, f'-2\, i\, f+2\, u\, \omega \right)+h_{x,\,{s}} \left(2\, m^2\, u^3\, f\, \ddot{V}+\omega\,  \left(i\, u\, f'-2\, i\, f+2\, u\, \omega \right)\right))\nonumber\\ 
&+h_{tt} \left(u \left(-u f''+4\, f'+2\, m^2\, u\,  \dot{V}\right)-12\, f+k^2\, u^2-2\, m^2\, V-2\, i\, u\, \omega +6\right)=0\\
&2 h_{ty} \left(u \left(f'+m^2\, u\,  \dot{V}\right)-3\, f-m^2\, V+3\right)\nonumber\\
&-u \left(u\, f\, h_{ty}''-2\, f\, h_{ty}'+i\, k\, u\, h_{tt}'+k\, u\, \omega\,  h_{x,\,{s}}+k\, u\, \omega\,  h_{x,\,{a}}-2\, i\, m^2\, u\, \omega\,  \delta \phi_y\,  \dot{V}+i\, u\, \omega\,  h_{ty}'\right)\nonumber\\
&+2\, i\, k\, u\, h_{tt}=0\\
&h_{x,\,{s}} \left(2\, u \left(f'+m^2\, u\,  \dot{V}\right)-6\, f+k^2\, u^2-2\, m^2\, V+4\, i\, u\, \omega +6\right)-u^2\, f'\, h_{x,\,{s}}'\nonumber\\
&-u^2\, f'\, h_{x,\,{a}}'+2\, u\, h_{x,\,{a}}\, f'-u^2\, f\, h_{x,\,{s}}''-u^2\, f\, h_{x,\,{a}}''+4\, u\, f\, h_{x,\,{s}}'+2\, u\, f\, h_{x,\,{a}}'-6\, f\, h_{x,\,{a}}\nonumber\\
&+k^2\, u^2\, h_{x,\,{a}}+2\, i\, k\, u\, h_{ty}+2\, m^2\, u^2\, h_{x,\,{a}}\,  \dot{V}-2\,m^2\, h_{x,\,{a}}\, V-2\, i\, u^2\, \omega\,  h_{x,\,{s}}'\nonumber\\
&-2\, i\, u^2\, \omega\,  h_{x,\,{a}}'
-2\, u\, h_{tt}'+6 h_{tt}(u)+2\, i\, u\, \omega\,  h_{x,\,{a}}+6 h_{x,\,{a}}=0\\
&h_{x,\,{s}} \left(2\, u \left(f'+m^2\, u\, \dot{V}\right)-6\, f+k^2\, u^2-2\,m^2\, V+4\, i\, u\, \omega +6\right)-u^2\, f'\, h_{x,\,{s}}'+u^2\, f'\, h_{x,\,{a}}'\nonumber\\
&-2\, u\, h_{x,\,{a}}\, f'-u^2\, f\, h_{x,\,{s}}''+u^2\, f\, h_{x,\,{a}}''+4\, u\, f\, h_{x,\,{s}}'-2\, u\, f h_{x,\,{a}}'
+6\, f\, h_{x,\,{a}}+k^2\, u^2\, h_{x,\,{a}}\nonumber\\
&-4\, i\, k\, m^2\, u^2\, \delta \phi_y\,  \dot{V}-2\, i\, k\, u^2\, h_{ty}'+6\, i\, k\, u\, h_{ty}-2\, m^2\, u^2\, h_{x,\,{a}}\,  \dot{V}+2\, m^2\, h_{x,\,{a}}\, V\nonumber\\
&-2\, i\, u^2\, \omega\,  h_{x,\,{s}}'+2\, i\, u^2\, \omega\,  h_{x,\,{a}}'-2\, u\, h_{tt}'+6\, h_{tt}-2\, i\, u\, \omega\,  h_{x,\,{a}}-6\, h_{x,\,{a}}=0\\
&-6\, h_{tt}+u\, (u\, f'\, h_{x,\,{s}}'-2\, f\, h_{x,\,{s}}'-2\, i\, k\, m^2\, u^3\, \delta \phi_y\, \ddot{V}+i\, k\, u\, h_{ty}'-2\, i\, k\, h_{ty}\nonumber\\
&+2\, m^2\, u^3\, h_{x,\,{s}}\, \ddot{V}-u\, h_{tt}''+4\, h_{tt}'+2\, i\, u\, \omega\,  h_{x,\,{s}}'-2\, i\, \omega\,  h_{x,\,{s}})=0\\
&k\, u \left(h_{x,\,{s}}'+h_{x,\,{a}}'\right)-i\, u\, \left(2\,m^2\, \delta \phi_y'\,  \dot{V}+h_{ty}''\right)+2\, i\, h_{ty}'=0\\
&h_{x,\,{s}}''=0,
\end{align}
where the following notations $ \dot{V}\equiv d V(X)/dX,\, \ddot{V} \equiv d^2 V(X)/dX^2$ are used.\\
The quasinormal modes are obtained using pseudo-spectral methods. For more details about the numerical procedure see \cite{Grieninger:2020wsb}.

\end{document}